\documentclass[aps,prl,reprint,superscriptaddress,10pt,bibnotes]{revtex4-2}
\usepackage{graphicx} 
\usepackage{hyperref}
\usepackage{amsmath}

\usepackage{verbatim}

\hypersetup{
    pdftitle={Spectral Diffusion Mitigation with a Laser Pulse Sequence},
}

\begin{document}

\title{Spectral Diffusion Mitigation with a Laser Pulse Sequence}

\newcommand{\affHU}{\affiliation{Department of Physics, Humboldt-Universität zu Berlin, Newtonstr. 15, 12489 Berlin, Germany}}
\newcommand{\affFBH}{\affiliation{Ferdinand-Braun-Institut, Gustav-Kirchhoff-Str. 4, 12489 Berlin, Germany}}

\author{Kilian Unterguggenberger}
\author{Alok Gokhale}
\affHU
\author{Aleksei Tsarapkin}
\affHU
\affFBH
\author{Wentao Zhang}
\affFBH
\author{Katja Höflich}
\affHU
\affFBH
\author{Herbert Fotso}
\affiliation{Department of Physics, University at Buffalo SUNY, Buffalo, New York 14260, USA}
\author{Tommaso Pregnolato}
\affHU
\affFBH
\author{Laura Orphal-Kobin}
\affHU
\author{Tim Schröder}
\email{tim.schroeder@physik.hu-berlin.de}
\affHU
\affFBH

\begin{abstract}
The optical spectrum of a quantum system is jointly determined by the properties of the emitter and the driving field. All-optical spectral control can hence be a promising method to engineer the properties of single photon emitters for quantum technological applications. It was proposed that driving a two-level system with a periodic sequence of optical $\pi$-pulses during the excited state lifetime shifts the emission and absorption maximum to an arbitrarily detuned pulse carrier frequency, enabling the mitigation of spectral diffusion in noisy emitters. In this article, we report on the first experimental observation of this effect. We implement the protocol on a solid-state emitter and reduce its inhomogeneously broadened optical linewidth close to the lifetime limit. By detuning the excitation laser, we are able to concentrate approximately half of the absorption to a freely selectable target frequency. Our approach is solely based on properties of coherently evolving quantum systems, rendering it applicable to a wide range of individual and ensembles of quantum emitters.
\end{abstract}

\maketitle

Some of the most intriguing fundamental phenomena of quantum optics can only be understood in the fully quantum-mechanical picture of light-matter interaction. Describing atom and field as a single coherently evolving quantum system reveals that the spectrum not only depends on the properties of the quantum emitter, but also on those of the driving field. A well-known historic example is the Mollow triplet: under strong, resonant continuous driving, the emission spectrum of a two-level system consists of a central peak shifted to the frequency of the driving field and two side peaks with a separation defined by the driving field strength \cite{mollowPowerSpectrumLight1969}. Almost ten years ago, Fotso \emph{et al.} predicted a similar emission spectrum outside the strong driving regime from applying optical $\pi$-pulses \cite{fotsoSuppressingSpectralDiffusion2016}. By flipping the sign of phase accumulation with each pulse, the average phase over the excited state lifetime is nullified. As a result, spontaneous emission occurs at the pulse carrier frequency, independent of its detuning to the emitter resonance.

In addition to the fundamental aspects of this scheme, lifting the strong driving requirement enables it to counteract spectral diffusion, a detrimental effect of electric field noise in solid-state emitters. Quantum dots \cite{vajnerQuantumCommunicationUsing2022}, color centers in diamond and other optically active solid-state spins at crystal impurities \cite{awschalomQuantumTechnologiesOptically2018} are compatible with photonic integration \cite{liScalableDeterministicIntegration2023, liHeterogeneousIntegrationSpin2024} and hence excellent candidate platforms for scalable quantum networks \cite{stasRobustMultiqubitQuantum2022, pompiliRealizationMultinodeQuantum2021, wehnerQuantumInternetVision2018, vajnerQuantumCommunicationUsing2022, awschalomQuantumTechnologiesOptically2018, kimblequantumInternet2008} and photonic quantum computing \cite{KokEtAlRMP2007, KnillLaflammeMilburn2001, huetDeterministicReconfigurableGraph2025}. However, efficient entanglement generation is presently limited by the inhomogeneous broadening of individual emitters and emitter ensembles in their dynamic electronic environment \cite{ambroseFluorescenceSpectroscopySpectral1991, kuhlmannChargeNoiseSpin2013, vuralCharacterizationSpectralDiffusion2020, orphal-kobinOpticallyCoherentNitrogenVacancy2023, pieplowQuantumElectrometerTimeresolved2025}. Unlike competing methods, all-optical elimination of the random detuning between emitter and driving field would not require external static electric fields \cite{duquennoyEnhancedControlSingleMolecule2024}, strain \cite{zhaiLargerangeFrequencyTuning2020} or feedback loops \cite{srockaDeterministicallyFabricatedStraintunable2020,brevoordLargerangeTuningStabilization2025}, but has not yet been achieved to date.

In this work, we report on the first experimental observation of spectral control with optical pulses as proposed by Fotso \emph{et al}. We apply the protocol to a single nitrogen-vacancy center (NV) in diamond and demonstrate a reduced inhomogeneous absorption linewidth near the lifetime limit, pinned to a target frequency selected by the driving field. Our approach does not make assumptions on the emitter beyond the universal two-level system, rendering it applicable to a wide range of atomic and solid-state quantum emitters. A simulation of the pulse-engineered absorption spectrum of a noisy two-level system supports our experimental findings. Our work establishes a novel phenomenon in driven quantum systems, with promising applications for scalable sources of indistinguishable single photons in quantum technologies.

\section{Simulated spectrum}

\begin{figure}
\centering
\includegraphics[width=\columnwidth]{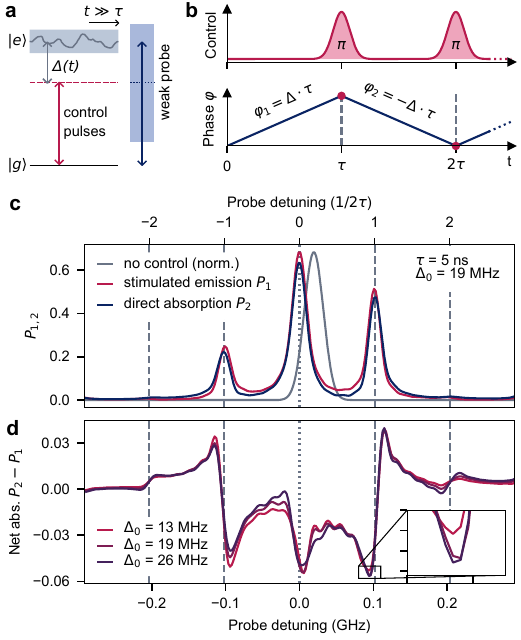}
\caption{Simulated light-matter interaction. \textbf{a}. Simplified level scheme. Spectral diffusion is indicated as a time-dependent transition energy from the ground state $|g\rangle$ to the excited state $|e\rangle$. \textbf{b}. Control sequence consisting of $N$ $\pi$-pulses with interpulse delay $\tau$ (top) and accumulation of relative phase (bottom). \textbf{c}. Stimulated emission and direct absorption spectra of a two-level system after a periodic sequence of $N = 12$ \(\pi\)-pulses. The absorption spectrum without control is a Gaussian with a FWHM of $2\sqrt{2 \ln{2}} \cdot \sigma_\Delta = 30\,\mathrm{MHz}$ centered at $\Delta_0$. \textbf{d}. Net absorption spectrum. The zoom inset highlights the growth of the closest satellite feature with increased detuning $\Delta_0$.}
\label{fig:theory}
\end{figure}

The spectral control protocol \cite{fotsoSuppressingSpectralDiffusion2016} consists of a periodic sequence of optical \(\pi\)-pulses with interpulse delay $\tau$ applied to a two-level system initialized in its excited state  (Fig. \ref{fig:theory}a). A \(\pi\)-pulse transfers the complete population of a two-level system from ground to excited state and vice versa. In even pulse intervals, the quantum system acquires a phase $\varphi_1 = \Delta \cdot \tau$ proportional to the detuning $\Delta$ from the driving field (Fig. \ref{fig:theory}b). In odd intervals, the two-level system is inverted by a \(\pi\)-pulse and acquires the same phase with opposite sign, $\varphi_2 = -\Delta \cdot \tau$. For an interpulse delay \(\tau < 1/\Gamma\) where \(\Gamma\) is the decay rate, approximately half of all photons obtain a net phase of $\varphi_1 + \varphi_2 = 0$ and are emitted at the pulse carrier frequency, independent from the detuning.

It was shown that the pulse sequence similarly controls the absorption spectrum of a two-level system \cite{fotsoAbsorptionSpectrumTwolevel2017}. The absorption is probed by an additional weak field that does not significantly affect the level populations. In the presence of charge noise-induced inhomogeneous broadening, the control protocol is expected to reduce the overall emission linewidth \cite{fotsoTuningSpectralProperties2023}. The time-dependent detuning $\Delta(t)$ with respect to the pulse carrier frequency is modeled by a Gaussian random distribution. During one iteration of the control sequence, the detuning is treated as static since spectral diffusion is slow compared to the excited state lifetime of an NV center \cite{orphal-kobinOpticallyCoherentNitrogenVacancy2023}. To match the experimental conditions in this work, we expand previous simulations and compute the absorption spectrum of an inhomogeneously broadened emitter (details in Supplementary Material \cite{supp}).

We first simulate the direct absorption and stimulated emission spectra individually (Fig.~\ref{fig:theory}c). They closely resemble the spectrum of the emitted photons simulated in reference \cite{fotsoTuningSpectralProperties2023}. The net absorption spectrum is then computed by taking the difference between the two (Fig.~\ref{fig:theory}d). Positive values indicate that direct absorption exceeds stimulated emission. We observe a dip at the pulse carrier frequency (dotted gray line) for all values of the central detuning $\Delta_0$, with a narrower linewidth compared to the uncontrolled case. The remaining spectral weight is located at satellite features with a frequency spacing $1/2\tau$ inversely proportional to the interpulse delay (dashed lines). As the central detuning $\Delta_0$ is increased, more spectral weight is shifted to the closest satellite features (inset). For a larger standard deviation $\sigma_\Delta$ of the detuning values, the features of the spectrum remain qualitatively unchanged \cite{supp}. If the interpulse delay is increased close to the lifetime, the spectral control becomes increasingly ineffective and this lineshape is destroyed (not shown).

\nocite{cohen_tannoudji_book1992, mollowStimulatedEmissionAbsorption1972, yurgensSpectrallyStableNitrogenvacancy2022, tsarapkinCleanYouNovel2026, binderQudiModularPython2017, batalovLowTemperatureStudies2009, mansonNitrogenvacancyCenterDiamond2006}

\begin{figure}
\centering
\includegraphics[width=\columnwidth]{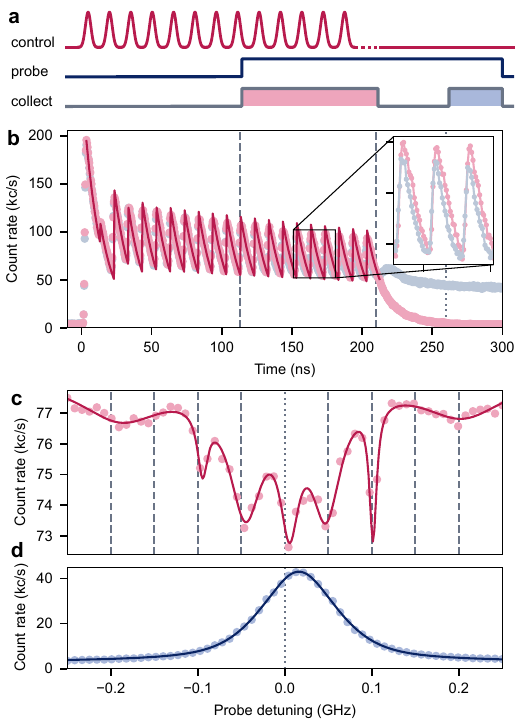}
\caption{Controlled absorption spectrum of a single NV center. \textbf{a}. Experimental sequence. The control field (red) consists of $N = 21$ periodic $\pi$-pulses. The probe field (blue) is turned on after 11 pulses. Collected photon detection events are integrated in time windows (light red and blue). \textbf{b}. Time-resolved fluorescence signal of the NV center for a probe detuning of zero (light red) and -0.25 GHz (light gray). A Master equation fit with an additional decay term is shown as a red line. The time windows are indicated by dashed and dotted lines. Here the interpulse delay $\tau = 10\,\mathrm{ns}$. \textbf{c}. Controlled spectrum fitted with a sum of seven Lorentzians. Dashed lines indicate expected satellite feature positions at $\pm n/2\tau$. \textbf{d}. Uncontrolled spectrum fitted with a pseudo-Voigt model. The dotted line indicates the pulse carrier frequency.}
\label{fig:absorption}
\end{figure}

\section{Implementation on an NV center}

We choose the NV center in diamond for the experimental realization of the spectral control protocol. Spectral instability is currently a key limiting factor of this solid-state emitter \cite{orphal-kobinOpticallyCoherentNitrogenVacancy2023}. To evaluate the spectral response to the control pulses, the pulse sequence in Fig.~\ref{fig:absorption}a is repeated many times for a range of probe frequencies (details in Supplementary Material \cite{supp}). First, the charge state of the NV is initialized. Then, the control pulse sequence starts and the first $\pi$-pulse prepares the NV in the excited state. The weak probe field turns on after a delay of a few control pulses. The fluorescence time traces for two exemplary probe field frequencies are shown in Fig.~\ref{fig:absorption}b. Here, the pulse carrier frequency is set near the mean emitter resonance. For a large probe detuning (red), the data is well described by a model based on the Master equation of the system \cite{fotsoSuppressingSpectralDiffusion2016}. We include an additional exponential decay term to account for pumping to a dark state beyond the two-level picture \cite{mansonNitrogenvacancyCenterDiamond2006}. We extract a decay time of $662\,\mathrm{ns}$ as the only free fit parameter. The lifetime of $1/\Gamma = 12.3\,\mathrm{ns}$ is measured independently \cite{supp}.

We obtain a spectrum with control pulses and one without control pulses from the same data set by averaging the fluorescence in separate time windows (dashed and dotted gray lines in Fig.~\ref{fig:absorption}b). Note that the relative spectroscopic signal is smaller during simultaneous driving by the $\pi$-pulses (inset) compared to when the system interacts only with the probe field. In our experiment, we scan the probe laser across the coherent zero-phonon line and collect the red-shifted phonon side-band (photoluminescence excitation spectroscopy). Direct absorption of the probe laser results in a spontaneously emitted photon and an increase in the collected signal. Stimulated emission induced by the probe laser is a coherent process; photon emission is redirected to the zero-phonon line and the collected signal is reduced.

The uncontrolled spectrum consists of a single direct absorption peak (Fig.~\ref{fig:absorption}d). We extract an uncontrolled linewidth of 104 MHz, corresponding to spectral broadening to eight times the lifetime limit of 12.9 MHz. By contrast, the controlled spectrum consists of a nearly symmetric array of dips that we attribute to stimulated emission (Fig.~\ref{fig:absorption}c). Remarkably, the central dip at the pulse carrier frequency has a width of only $27\,\mathrm{MHz}$, corresponding to only twice the lifetime limit. This strong refocusing to the target frequency indicates that the stochastically distributed detuning could be canceled for a significant portion of the spectral weight. In agreement with the theoretical prediction, the central dip is accompanied by satellite dips to either side at the simulated regular intervals, with decreasing amplitude. Since the pulse carrier frequency is not perfectly centered on the uncontrolled resonance, the dip amplitudes are asymmetric. All dips are significantly narrower than the uncontrolled resonance. As absorption and emission are closely related on a fundamental level, we can expect similarly narrow linewidths for the emission spectrum. For reference, we also run the identical sequence with a pulse rotation angle different from \(\pi\) and observe that the refocusing effect vanishes (see Supplementary Material \cite{supp}).

\begin{figure}
\centering
\includegraphics[width=\columnwidth]{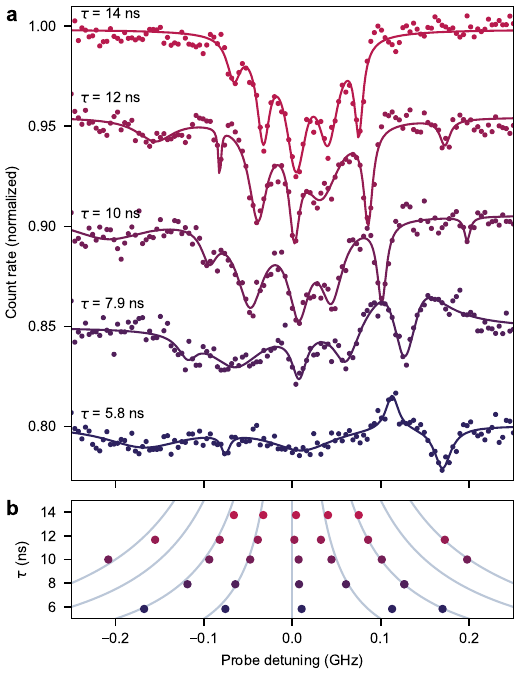}
\caption{Effect of the interpulse delay. \textbf{a}. Controlled spectra for different interpulse delays \(\tau\). The fits are sums of five to seven Lorentzians. Here, the count rate is normalized to the outermost data points. Individual spectra are offset by 0.05 for visual clarity. \textbf{b}. Satellite feature positions extracted from the fits. The simulated positions of $\pm n/2\tau$ are shown as gray lines.}
\label{fig:interpulse-delay}
\end{figure}

In the following, we investigate how the control parameters change the spectrum. Fig.~\ref{fig:interpulse-delay} shows the effect of the interpulse delay. As can be clearly seen in the spectra, the satellite features are indeed spaced further apart with decreasing interpulse delay, following the predicted $1/2\tau$ relationship. For shorter interpulse delay, direct absorption peaks emerge as well. In good agreement with the theory, direct absorption can exceed stimulated emission for parts of the spectrum. Note that for spectral control, an interpulse delay $\tau < 1 / (\pi\cdot\Delta)$ is optimal since it is predicted to maximize the spectral weight at the target frequency \cite{fotsoSuppressingSpectralDiffusion2016}. For one standard deviation of the inhomogeneous linewidth, this translates to $\tau < 6.1\,\mathrm{ns}$. In our measurement however, a higher excitation rate by the control pulses also reduces the measurement contrast, preventing us from observing this effect.

\begin{figure}[t]
\centering
\includegraphics[width=\columnwidth]{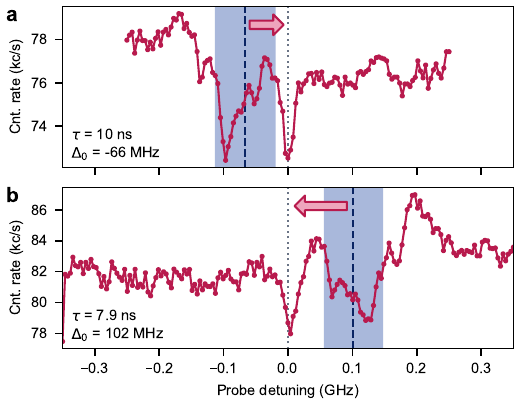}
\caption{Detuned pulse carrier frequency. \textbf{a}-\textbf{b}. Controlled spectra with different detunings \(\Delta_0\) and interpulse delays \(\tau\) are shown (red). The uncontrolled resonance is shown in blue (centered at the dashed line, FWHM represented by shaded region). A red arrow indicates the spectral shift achieved by the control pulses to the pulse carrier frequency (dotted gray line).}
\label{fig:off-resonance}
\end{figure}

Moreover, we introduce a deliberate detuning \(\Delta_0\) between the pulse carrier frequency and the center of the uncontrolled emitter resonance. Fig.~\ref{fig:off-resonance}a and b show the spectrum for a negative detuning of five natural linewidths and a positive detuning of eight natural linewidths, respectively. Strikingly, even for these large mean detunings, a dip at the pulse carrier frequency stands out from the data. Our observation confirms that a refocusing of about half of the absorption to the pulse carrier frequency is possible. In agreement with our simulation, the relative spectral weight at the target frequency decreases with increased detuning, while the satellite dip closest to the uncontrolled resonance grows. The experimentally obtained spectra explicitly demonstrate that the control protocol is capable of shifting the resonance of a two-level system in an arbitrary direction.

\begin{figure}
\centering
\includegraphics[width=\columnwidth]{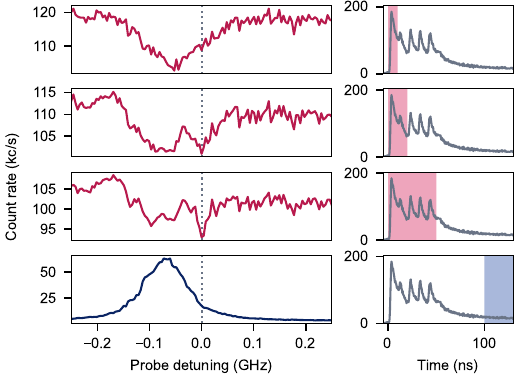}
\caption{Temporal evolution of the spectrum. Four spectra (left column) and the corresponding time window (right column, shaded region) are shown. The last spectrum shows the uncontrolled linewidth (blue). The pulse carrier frequency is indicated by the dotted gray line. Here, the interpulse delay $\tau = 10\,\mathrm{ns}$ and the number of control pulses $N = 5$.}
\label{fig:evolution}
\end{figure}

Finally, we investigate how the spectrum evolves over time to determine the number of pulses required to reach the controlled state (Fig.~\ref{fig:evolution}). In contrast to the measurements above, the probe field now starts immediately with the first control pulse. To obtain the time evolution, we vary the width of the time window for the controlled spectrum from one to five pulses. After the first \(\pi\)-pulse, we observe a broad probe field-induced fluorescence reduction at the uncontrolled emitter resonance. A possible explanation is that the probe field contributes to the population transfer of the \(\pi\)-pulse, resulting in a reduced excited state population. After just two pulses, a narrow dip emerges at the pulse carrier frequency, indicating the onset of detuning cancellation. This observation is in excellent agreement with the theory. After initialization into the excited state by the first pulse, the two-level system is only inverted once by the second pulse. If a photon is emitted just before the third pulse, the system has acquired the same phase once with positive and once with negative sign, and hence the effective phase at emission is zero.

\section{Conclusion and outlook}\label{discussion}

In conclusion, we have demonstrated that optical pulses can refocus the resonance of a noisy quantum emitter onto a tunable target frequency. Specifically, we reduced the inhomogeneous linewidth by a factor of four and shifted a significant part of the absorption to a target frequency detuned by up to eight natural linewidths. The control protocol is applicable to every two-level system as long as driving with two \(\pi\)-pulses within one lifetime is experimentally feasible. Our application to an NV center in diamond uses only commercially available optical switching and electronic driving technology. Compared to other spectral engineering methods, such as frequency modulation of solid-state emitters \cite{lukinSpectrallyReconfigurableQuantum2020}, our approach does not require nano-fabrication of electrical gates or means to achieve mechanical stress. Notably, even large ensembles of dissimilar quantum emitters can in principle be brought into resonance with each other using the control protocol. Especially for ensembles, conventional feedback loops would require a larger experimental overhead. Future work could address a direct measurement of the emission linewidth and photon indistinguishability \cite{fotsoPulseenhancedTwophotonInterference2019} in presence of the optical control pulses, providing further insight into this novel mechanism of quantum optical systems.


\bigskip

\emph{Acknowledgments}—We thank Viviana Villafañe for sample annealing. Ion implantation was carried out at the Ion Beam Center at the Helmholtz-Zentrum Dresden-Rossendorf e.~V., a member of the Helmholtz Association, where we thank Shavkat Akhmadaliev for assistance.
Funding for this project was provided by the German Federal Ministry of Research, Technology and Space (BMFTR, projects  no. 13N14921 and QPIS no. 16KISQ032K) as well as by the European Research Council within an ERC Starting Grant (project QUREP no. 851810). We acknowledge further funding from Berlin Quantum. Herbert Fotso is supported by the National Science Foundation under Grant number NSF PHY-2014023.

\bigskip

\emph{Author contributions}—K.U.: Investigation (lead), Formal Analysis (lead), Methodology (equal), Writing - Original Draft (lead). A.G.: Investigation (supporting), Methodology (equal). A.T., W.Z., K.H. and T.P.: Methodology (equal). H.F.: Formal Analysis (supporting), Methodology (equal). L.O.K.: Conceptualization (supporting), Methodology (equal). T.S.: Conceptualization (lead). All authors: Writing - Review and Editing (equal).

\bigskip

\emph{Data availability}—The measurement data that support the findings of this study will be made publicly available upon publication of the peer-reviewed version of this manuscript.

\bibliography{bibliography}

\end{document}


\title{Supplementary Material: Spectral Diffusion Mitigation with a Laser Pulse Sequence}

\newcommand{\affHU}{\affiliation{Department of Physics, Humboldt-Universität zu Berlin, Newtonstr. 15, 12489 Berlin, Germany}}
\newcommand{\affFBH}{\affiliation{Ferdinand-Braun-Institut, Gustav-Kirchhoff-Str. 4, 12489 Berlin, Germany}}

\author{Kilian Unterguggenberger}
\author{Alok Gokhale}
\affHU
\author{Aleksei Tsarapkin}
\affHU
\affFBH
\author{Wentao Zhang}
\affFBH
\author{Katja Höflich}
\affHU
\affFBH
\author{Herbert Fotso}
\affiliation{Department of Physics, University at Buffalo SUNY, Buffalo, New York 14260, USA}
\author{Tommaso Pregnolato}
\affHU
\affFBH
\author{Laura Orphal-Kobin}
\affHU
\author{Tim Schröder}
\email{tim.schroeder@physik.hu-berlin.de}
\affHU
\affFBH

\maketitle

\renewcommand{\theequation}{S\arabic{equation}}
\renewcommand{\thefigure}{S\arabic{figure}}

\section{Simulation}

We simulate the absorption spectrum of a noisy two-level system under a periodic sequence of Gaussian \(\pi\)-pulses. The system of \textit{two-level system + photon bath + control field} is described, in the rotating wave approximation by the Hamiltonian:
\begin{equation}
    H = \sum_{k} \omega_k a^{\dagger}_{k}a_{k} + \frac{\Delta(t)}{2} \sigma_z - i \sum_{k} g_{k} \left( a^{\dagger}_{k} \sigma_- - a_{k} \sigma_+ \right) + \frac{\Omega_x(t)}{2}(\sigma_+ + \sigma_-). 
\label{eq:hamiltonian_1}
\end{equation}
Where $\sigma_z=|e\rangle\langle e|-|g\rangle\langle g|$, $\sigma_+ =|e\rangle\langle g|$, and $\sigma_- = |g\rangle\langle e| = (\sigma_+)^\dagger$ are respectively, the $z$-axis Pauli matrix, the raising, and the lowering operators for the two-level system. $a_k$ ($a^{\dagger}_k$) is the annihilation (creation) operator of the $k$-th photon mode, $g_k$ is its coupling strength to the emitter, and $\omega_k$ is the detuning from $\omega_0$ of mode $k$.  We consider pulses such that $\Omega_x(t)$ is of a Gaussian shape during the time $t_{\pi}$ such that the area under the $\Omega_x \;\; VS \;\; t$ curve during this time is equal to $\pi$ and $\Omega_x(t) = 0$ otherwise. $\Delta(t)=\omega_1(t)-\omega_0$ is the time-dependent detuning of the two-level system's transition frequency from the pulse carrier frequency. This detuning is considered to vary randomly in time following a random Gaussian distribution centered around $\Delta_0$ and with standard deviation $\sigma_{\Delta}$.

In the simulation, we assume that the system is initially prepared in the excited state. Although this condition is easily relaxed, since if the system is initially in its ground state, it is driven to its excited state by the first $\pi$ pulse and our condition amounts to obtaining the absorption spectrum one pulse earlier, which does not affect the lineshape. 
In the absence of all control ($\Omega_x(t) = 0$ for all times), spontaneous decay will occur, and for a static detuning, the corresponding emission rate is $\Gamma = 2\pi\int g_k^2 \; \delta(\omega_k - \Delta) \; dk$.

The density matrix operator of the two-level system can be written as:
\begin{equation}
 \rho(t) = \rho_{ee}(t) |e\rangle \langle e| + \rho_{eg}(t) |e\rangle \langle g|
 + \rho_{ge}(t) |g \rangle \langle e|  + \rho_{gg}(t) |g\rangle \langle g| \; ,
\label{eq:TrueDensMatr}
\end{equation} 
The master equation governing its time-evolution can be obtained through the independent rate approximation by independently adding up, in the time-evolution of the matrix elements of $\rho$, terms due to the radiation bath, to the incident field, and the damping terms responsible for spontaneous emission~\cite{Cohen_Tannoudji_Book1992}.
This master equation reads:
\begin{equation}
\label{eq:MasterEquation_1} 
\begin{split}
 \frac{d \rho_{ee}}{dt} &= i \frac{\Omega_x(t)}{2}(\rho_{eg} - \rho_{ge}) - \Gamma \rho_{ee} \; ,   \\
\frac{d \rho_{gg}}{dt} &= -i \frac{\Omega_x(t)}{2}(\rho_{eg} - \rho_{ge}) + \Gamma \rho_{ee} \; ,  \\
\frac{d \rho_{ge}}{dt} &= (i\Delta(t) -\frac{\Gamma}{2}) \rho_{ge} -i \frac{\Omega_x(t)}{2}\left(\rho_{ee} - \rho_{gg}\right) \; , \\
\frac{d \rho_{eg}}{dt} &= (-i\Delta(t) -\frac{\Gamma}{2}) \rho_{eg} +i \frac{\Omega_x(t)}{2}\left(\rho_{ee} - \rho_{gg}\right) .
\end{split}
\end{equation}
The absorption spectrum after some time $T$ can be written as:
\begin{equation}
  Q(\omega) = P_2(\omega) - P_1(\omega).
\end{equation}
Where $P_1(\omega)= A \times \mathrm{Re} \left\{ \mathcal{P}_1(\omega) \right\} $  and $P_2(\omega)= A \times \mathrm{Re} \left\{ \mathcal{P}_2(\omega) \right\}$ 
with
\begin{equation}
 \mathcal{P}_2(\omega) =  \int_0^T \; dt\; \int_0^{T-t} d\theta \; \langle  \sigma_-(t) \sigma_+(t+\theta) \rangle \mathrm{e}^{-i \omega \theta}
\end{equation}
and
\begin{equation}
 \mathcal{P}_1(\omega) =  \int_0^T \; dt\; \int_0^{T-t} d\theta \; \langle \sigma_+(t+\theta) \sigma_-(t) \rangle \mathrm{e}^{-i \omega \theta}.
\end{equation}
$A$ is a proportionality constant that does not affect the spectral lineshape. $P_1(\omega)$ can be viewed as the stimulated emission spectrum of the two-level system and $P_2(\omega)$ as the direct absorption so that the difference yields the net absorption~\cite{mollowPowerSpectrumLight1969}. Both quantities can be evaluated separately and the absorption spectrum obtained by taking the difference.

\begin{figure}
    \centering
    \includegraphics[width=\textwidth]{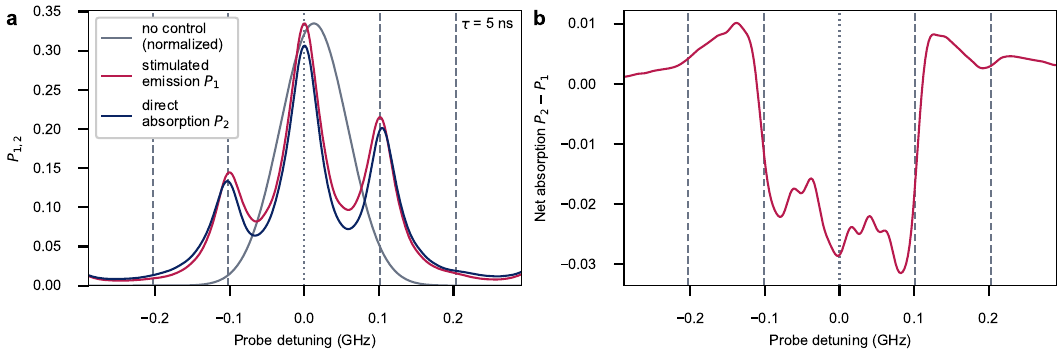}
    \caption{Simulated spectrum for strong inhomogeneous broadening. \textbf{a}. Stimulated emission $P_1$ (red) and direct absorption $P_2$ (blue) in comparison to the normalized uncontrolled spectrum (gray). Here, the inhomogeneous linewidth without control is $107\,\mathrm{MHz}$, matching the experimentally observed linewidth. The interpulse delay is $5\,\mathrm{ns}$. \textbf{b}. Net absorption spectrum obtained by evaluating the difference $P_2 - P_1$. The dotted line indicates the pulse carrier frequency.}
    \label{fig:theory-broad}
\end{figure}

To evaluate $P_1(\omega)$ and $P_2(\omega)$, we numerically integrate the master equation (\ref{eq:MasterEquation_1})  on a discretized time axis following the procedure described in Ref [\onlinecite{fotsoAbsorptionSpectrumTwolevel2017}] with $\Omega_x(t)$ defined by our periodic sequence of Gaussian pulses, and $\Delta(t)$ given by the above described random Gaussian distribution. The system is initialized in the excited state and we calculate the time-evolved matrix elements of the density matrix. By evaluating the relevant correlation functions, we obtain the direct absorption spectrum and the stimulated emission and then, the net absorption spectrum by taking the difference  \cite{fotsoAbsorptionSpectrumTwolevel2017, mollowPowerSpectrumLight1969, mollowStimulatedEmissionAbsorption1972}. The Gaussian pulses have a duration $t_\pi = 1.6\,\mathrm{ns}$ to match the experiment. All parameters are converted to laboratory units using the measured NV lifetime. In addition to the data in the main text, a spectrum simulated for the experimentally observed spectral diffusion is shown in Fig. \ref{fig:theory-broad}.

\section{Sample}

The diamond is a $2\times2\times0.5\,\mathrm{mm}^3$ single crystal grown by chemical vapor deposition with a nitrogen concentration of less than $5\,\mathrm{ppb}$ (Element Six ELSC). We create vacancies by implanting the sample with $^{12} \mathrm{C}$ ions (energy $11\,\mathrm{MeV}$, dose $10^{9}\,\mathrm{ions/cm}^2$, angle $7^\circ$) \cite{yurgensSpectrallyStableNitrogenvacancy2022}. We simulate a depth of $4.9 \pm 0.2\,\mathrm{um}$ for the implantation layer with SRIM. Then, we anneal the sample in vacuum ($10^{-7}\,\mathrm{hPa}$) at $1050\,\mathrm{^\circ C}$ for 16 hours to enable vacancy diffusion and combination with nitrogen atoms. After annealing, the sample is cleaned in a 1:1:1 boiling mixture of sulfuric, perchloric and nitric acid to remove graphite from the surface. We then characterize the sample in the confocal microscope and measure an NV density of $10^{7}\,\mathrm{cm}^{-2}$ inside the implanted layer.
To overcome the limited photon collection efficiency from NV centers in bulk diamond, we mill hemispherical solid-immersion lenses with a focused ion beam \cite{tsarapkinCleanYouNovel2026}. By setting the lens radius to the simulated implantation depth, we ensure an optimal positioning of the emitter inside the lens in one dimension. We mill an array of lenses and post-select one lens with a bright NV nearly centered in the other two dimensions.

\section{Measurement setup}

\begin{figure}
\centering
\includegraphics[]{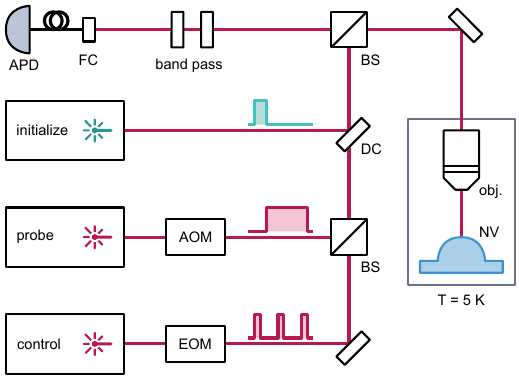}
\caption{Optical measurement setup. Simplified sketch showing the excitation and collection paths of the experimental setup. The two resonant lasers (control and probe) are pulsed with an EOM and AOM, respectively, and overlayed on a beam splitter (BS). A dichroic mirror (DC) combines them with the initialization laser. A beam splitter (10:90 R:T) separates excitation and collection before the cyrostat. Edgepass filters in the collection path are combined to form a bandpass filter for 650 to $800\,\mathrm{nm}$ before coupling into a single mode fiber (FC).}
\label{fig:setup}
\end{figure}

We use a home-built confocal microscope to excite and collect fluorescence from a single NV center. A sketch of the optical setup is shown in Fig. \ref{fig:setup}. The sample is cooled to $5\,\mathrm{K}$ in a closed-cycle helium cryostat to suppress phonons (Attocube attoDRY800). Galvo mirrors and a 4f-system allow us to scan across the sample in two dimensions. The objective (Attocube LT-APO VISIR) is mounted inside the cryostat on a piezo stage that moves along the optical axis to scan the focus in depth. We spectrally filter for phonon side-band fluorescence with edgepass filters (Thorlabs) before collection into a single-mode fiber (Thorlabs) connected to an avalanche photo diode (APD, Excelitas). The facet of the collection fiber serves as the pinhole of the confocal microscope.

\begin{figure}[p]
    \centering
    \includegraphics[width=\textwidth]{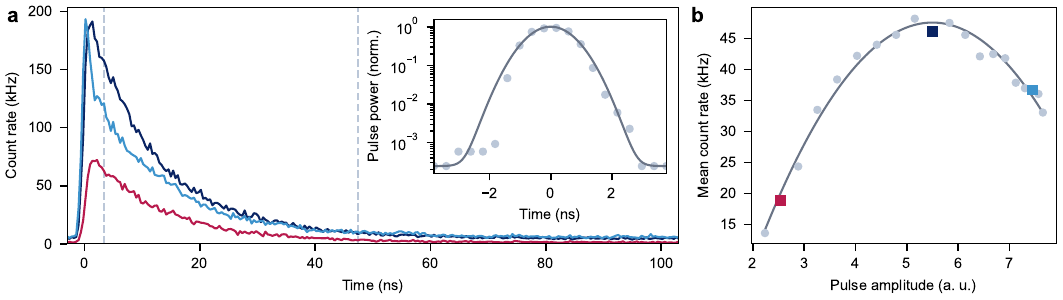}
    \caption{$\pi$-pulse calibration. \textbf{a}. Fluorescence response of NV to single excitation pulse with varying amplitude. The mean count rate in a time window (dashed lines) is representative for the excited state population. Inset: Excitation pulse measured via back-reflection off the sample. The pulse is Gaussian with a FWHM of $1.6\,\mathrm{ns}$. \textbf{b}. Mean count rate with respect to pulse amplitude. The data is fit by a Gaussian function. The center of the fit is the power required for a $\pi$-pulse.}
    \label{fig:calibration}
\end{figure}

For resonant excitation, two tunable external cavity diode lasers at $637\,\mathrm{nm}$ are used (Toptica DL pro HP). Their frequencies are locked continuously to a wavemeter (HighFinesse WS7). In the control laser path, Gaussian pulses of fixed length (Fig. \ref{fig:calibration}a, inset) are carved out from the continuous-wave laser with a fiber-coupled electro-optic amplitude modulator (EOM, Jenoptik AM635b). A modulator bias controller (OZ Optics) with an integrated photodiode locks the EOM to the minimal transmission bias point. To stabilize the optical control pulse area, we use a half-wave plate on a motorized rotation mount in combination with a polarizer and a photodiode. Since the pulses are too fast to be registered individually by this system, we stabilize the average power over the known sequence. Amplitude modulation and power stabilization of the probe laser is realized with a fiber-coupled AOM (Gooch \& Housego). Charge state initialization is performed with a $525\,\mathrm{nm}$ diode laser (LABS electronics). The polarization of each laser is optimized individually to the dipole axis of the NV transition with a half-wave plate.

Photon detection events are registered by a time tagger (quTools). An arbitrary waveform generator (Zurich Instruments HDAWG) drives the control laser EOM, and triggers the probe laser AOM driver, the initialization laser and the time tagger. The experiment is controlled by the Python-based modular control suite Qudi \cite{binderQudiModularPython2017}.

Due to the small signal-to-background ratio in the presence of the control pulses, the measurement is exceptionally sensitive against fluctuations. Long integration times of several hours per spectrum, repeated frequency scans and the aforementioned continuous stabilization of all optical powers and frequencies were therefore required to achieve reliable measurement outcomes.

\begin{figure}[p]
    \centering
    \includegraphics[width=\textwidth]{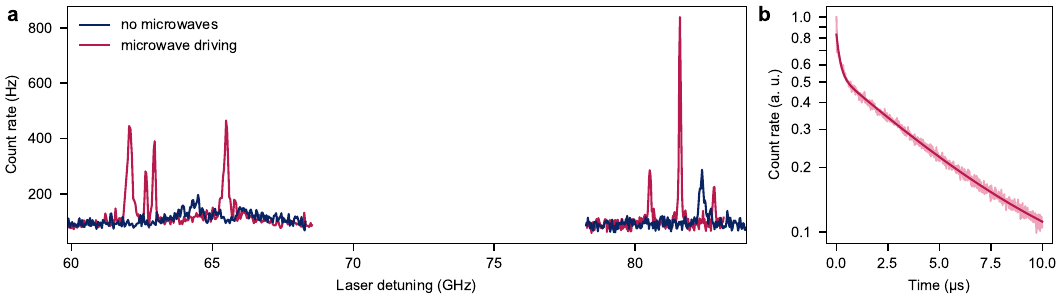}
    \caption{The NV as a two-level system. \textbf{a}. PLE spectra of the single NV center under study in this work. Data with (without) microwave driving is shown in red (blue). \textbf{b}. Fluorescence decay of the $|S_z\rangle \rightarrow|E_x, S_z\rangle$ transition under resonant excitation. The data is fitted with a bi-exponential decay.}
    \label{fig:tls-decay}
\end{figure}

\begin{figure}
\centering
\includegraphics[]{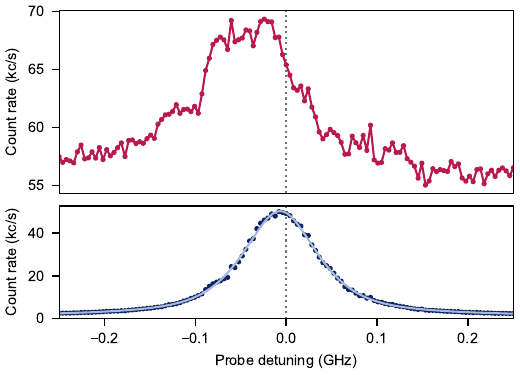}
\caption{Reference measurement with $\pi/2$-pulses. The top (bottom) plot shows the spectrum in the time window with (without) control pulses. Here, the interpulse delay $\tau = 10\,\mathrm{ns}$.}
\label{fig:not-pi}
\end{figure}

\clearpage

\section{Emitter characterization}

Fig. \ref{fig:tls-decay}a shows a large-range PLE spectrum of the selected NV center. Under continuous resonant microwave driving of the ground state splitting frequency (2.87 GHz), eight resonances become visible (red data). Following \cite{batalovLowTemperatureStudies2009}, we assign the lower energy resonances to three spin-conserving and two fully allowed spin-flipping transitions to the $E_y$ excited state manifold. The three higher energy resonances all stem from spin-conserving transitions to the $E_x$ excited state manifold. In this work, we operate at the brightest resonance, corresponding to the cycling transition $|S_z\rangle \rightarrow|E_x, S_z\rangle$. It is the only resonance that is visible without microwave driving (blue data).  We attribute the resonance shift to heating by the microwaves. To characterize the cyclicity of the transition, we measure the fluorescence decay under continuous resonant excitation (Fig. \ref{fig:tls-decay}b). We observe a bi-exponential decay with time constants of $180\,\mathrm{ns}$ and $4.8\,\mathrm{us}$, respectively. Following \cite{mansonNitrogenvacancyCenterDiamond2006}, we assign the faster process to decay to the metastable singlet state, and the slower process to decay to the $|S_{x,y}\rangle$ spin projections. We conclude that the cyclicity on the time scales of our experiment is sufficient for a description of the NV as a two-level system.

\section{Pulse calibration}

In Fig. \ref{fig:calibration}a, the fluorescence response to a single resonant excitation pulse is shown for different pulse amplitudes. An exponential fit to the data yields an excited state lifetime of $12.3\,\mathrm{ns}$. The pulse amplitude corresponding to a $\pi$-rotation is found by fitting the mean fluorescence in a time window (dashed lines) with a Gaussian (\ref{fig:calibration}b).

\section{Spectrum for pi/2 pulses}

Fig. \ref{fig:not-pi} shows the spectrum with an altered pulse sequence composed of $\pi/2$-rotations instead of $\pi$-rotations. There is no feature at the pulse carrier frequency in the spectrum with control pulses (gray line in top plot). Here, the interpulse delay $\tau = 10\,\mathrm{ns}$. The spectrum without control pulses shows the same Pseudo-Voigt shape as in the other measurements.

\bibliography{bibliography}